# Reply to the second version of the comment on "Influence of the Dzyaloshinskii-Moriya Exchange Interaction on Quantum Phase Interference of Spins"

W. Wernsdorfer[1], T.C. Stamatatos[2], and G. Christou[2]

[1]Institut Néel, CNRS & Université J. Fourier, BP 166, 38042 Grenoble Cedex 9, France
[2]Department of Chemistry, University of Florida, Gainesville, Florida 32611-7200, USA

Resonant quantum tunneling has first been studied in $Mn_{12}$ molecular wheels in [1] and quantum phase interference in [2]. However, we showed in [3] that the data published in [2] are not consistent and that the Landau-Zener method has been applied outside its range of applicability. It is important to note for our studies in [4] that our molecule [1] exhibits hysteresis loops with resonant tunneling steps that are much more narrow than those of [2], allowing us for the first time to study in detail the quantum effects involved. In particular, we showed in [4] that the quantum phase interference is strongly dependent on the direction of the Dzyaloshinskii-Moriya (DM) vector.

We are now happy to see that the authors of the recently submitted comment [5] on our letter [4] are now agreeing [6] with most of the issues that we first explained in [1,3,4,7,8]. We summarize here the main issues.

We pointed out [3,4,7,8] that DM interactions result in general from pairwise interactions of neighboring spins that do not have an inversion center. This condition is fulfilled most of the time in SMMs even when the entire molecule has an inversion center. In particular, for our $Mn_{12}$ molecular wheel with 12 pairwise DM interactions, it is extremely unlikely that all these individual DM contributions would cancel out for all applied fields, and all possible excitations and dynamics. The possibility of a DM interaction in the $Mn_{12}$ molecular wheel is thus well justified. A related example is the Kagomé spin lattice, which has also an inversion center: Nevertheless, there are many studies of DM interactions in the Kagomé spin lattice [9]. We should also repeat that our $Mn_{12}$ wheels crystallize in a very low symmetry space group and that the magnetic anisotropy axes are tilted with respect to the geometrical axes (see Fig. 2 of [3]). In particular, the individual DM vectors (from pairwise interactions of neighboring spins) are not aligned with the magnetic anisotropy axes. Considering in addition that tiny structural distortions will also break symmetries, it is clear to us that DM interactions have to be present in the $Mn_{12}$mda molecule, that the parity argument in [6] cannot be applied here, and that there is no particular reason why the "effective" DM vector of our model should be aligned with the easy axis of magnetization. **Note that a tiny**

**random symmetry-breaking defect is sufficient to invalidate the parity argument in [6], whereas the main DM effect comes from the systematically tilted DM vectors.**

We are also happy to see that the authors of [5] have improved their numerical program for calculating tunnel splittings [6]. We pointed out in [7] that because the DM interaction can shift the positions of tunnel splittings, a "search procedure" is needed to find the true tunnel splittings, that is, the tunnel splitting has to be taken at the minimum distance between the corresponding eigenvalues. This is not so simple because, for example, the hard magnetization axis turns slightly away from the x-axis and the $k = 0$ resonance shifts away from zero field (depending on the direction of the DM vector). When done properly, one should find the findings stated in our Phys. Rev. Lett. [4]. In particular, a small DM interaction does not quench the quantum phase interference. The calculation in Fig. 1 of [5] is completely meaningless because the tunnel splitting was not taken at the minimum distance between the corresponding eigenvalues. The calculation in Fig. 1 of the new version of the comment [6] seems correct. It leads the authors to conclude that the "transverse projections of the DM vector and the transverse single-ion anisotropies are incommensurate, forcing different minima in $k = 1(A)$ to occur at different transverse field orientations". The authors therefore suggest that this effect could be behind the rounding of the $k = 1(A)$ Berry phase minima observed in [2,4]. However, we would like to point out that the data in [2] were measured outside the Landau-Zener regime [3] and the data in [4] were measured under optimum conditions, that is, we experimentally turned the fields to observe maximum oscillation. As explained in [4], we could not reach the fast field sweep rate regime and could therefore not obtain the tunnel splitting, which probably led to the main part of the "rounding". In addition, tiny structural distortions will also lead to a "rounding".

In order to explain the observed tunnel resonances and tunnel probabilities, and to study the influence of DM interaction on quantum phase interference, we modeled the 12-spin-system with a simple dimer model of two ferromagnetically coupled spins $S_1 = S_2 = 7/2$ [1,4]. Although this model can be questioned [3], it represents a useful simplification that keeps the required calculations manageable, has been found to describe well the lowest energy levels, and allowed us a qualitative discussion of the observed tunnel rates. The simple model employed does not affect the generality of the obtained conclusions about the influence of the DM interaction [4].

We would like to conclude by repeating one of the last sentences of our paper [4]: The deviations between our data and the employed model *"should motivate more theoretical work*

*on the subject, as well as extensions to more sophisticated models for the Mn$_{12}$ wheel involving two sets of six independent Mn spins."*